\documentclass[11pt]{revtex4-1}

\usepackage{graphicx}

\begin{document}

\title{Particle creation and non-equilibrium thermodynamical prescription of dark fluids for universe bounded by an event horizon}

\author{Subhajit Saha\footnote {subhajit1729@gmail.com}}
\affiliation{Department of Mathematics, Jadavpur University, Kolkata 700032, West Bengal, India}

\author{Atreyee Biswas\footnote {atreyee11@gmail.com}}
\affiliation{Department of Natural Sciences, West Bengal University of Technology, BF-142, Sector-1, Saltlake, Kolkata 700064, West Bengal, India}

\author{Subenoy Chakraborty\footnote {schakraborty.math@gmail.com}}
\affiliation{Department of Mathematics, Jadavpur University, Kolkata 700032, West Bengal, India.}

\begin{abstract}

In the present work, flat FRW model of the universe is considered to be an isolated open thermodynamical system where non-equilibrium prescription has been studied using the mechanism of particle creation. In the perspective of recent observational evidences, the matter distribution in the universe is assumed to be dominated by dark matter and dark energy. The dark matter is chosen as dust while for dark energy, the following choices are considered: (i) Perfect fluid with constant equation of state and (ii) Holographic dark energy. In both the cases, the validity of generalized second law of thermodynamics (GSLT) which states that the total entropy of the fluid as well as that of the horizon should not decrease with the evolution of the universe, has been examined graphically for universe bounded by the event horizon. It is found that GSLT holds in both the cases with some restrictions on the interacting coupling parameter.

\end{abstract}

\maketitle

\section{Introduction}	

Cosmological models using irreversible thermodynamics along with Einstein's general relativity (GR) have recently gained great interest among cosmologists. With this new perspective, it is possible to analyze different cosmological models which have blockages in equilibrium thermodynamic set-up. Eling, Guedens and Jacobson \cite{Eling1} first introduced irreversible thermodynamics in cosmology after he failed to reproduce Einstein's field equations from the first law of thermodynamics in $f(R)$-gravity and concluded that a non-equilibrium thermodynamic treatment is essential for curvature correction to entropy. They added an extra entropy term $d_{i}S$ called entropy production term in the entropy balance equation $dS=\frac{dQ}{T}+d_{i}S$, where they have argued $d_{i}S$ as the bulk viscosity production term which can be determined by imposing energy-momentum conservation. In general, the entropy balance relation in non equilibrium thermodynamics is of the form $dS=d_{e}S+d_{i}S$, where $d_{e}S$ is the rate of entropy exchange with the surroundings while  $d_{i}S$ ($\geq 0$) comes from the process occuring inside the system. In particular, $d_{i}S$ is zero for a reversible process and positive for an irreversible process. In cosmology, $d_{i}S$ has no clear interpretation as it depends on the internal production process. This term has been identified as a dissipative effect by exploiting a fluid dynamics description of the local causal horizon kinematics while considering irreversible thermodynamics. Subsequently, a lot of work have been done in order to interpret the entropy production term in various ways. In 2009, G. Chirco and S. Liberati \cite{Chirco1}, have shown that the dissipative character leading to non-equilibrium spacetime thermodynamics is actually related (both in GR as well as in $f(R)-$gravity) to non-local heat fluxes associated with the purely gravitational/internal degrees of freedom of the theory. In particular, in the case of GR they showed that the internal entropy production term is identical to the so called tidal heating term of Hawking-Hartle \cite{Hawking1}. Similarly, for the case of $f(R)$-gravity, the dissipative effects can be associated with the generalization of this term plus a scalar contribution whose presence is clearly justified within the scalar-tensor representation of the theory.

Two most profound mysterious components of matter present in the universe are dark energy (DE) and dark matter (DM). According to recent observational data obtained from Type Ia Supernova, our Universe is now experiencing an accelerated expansion. This late time cosmic acceleration is interpreted by the existence of DE which is believed to have a huge negative pressure and therefore it is gravitationally repulsive. A common and natural candidate for DE is the cosmological constant but it is not widely used in the literature due to its limitations and drawbacks. Several alternative dynamical DE models have been proposed but the nature of DE (except that it has a huge negative pressure) is still a mystery. On the other hand, the behaviour of galatic rotation curves and the mass discrepancy in the cluster of galaxies are interpreted by the existence of DM which unlike DE, is gravitationally attractive in nature. As we know that DE is homogeneously distributed while DM is inhomogenously distributed, occuring in clumps around ordinary matter. Interaction between these two dark components is expected to be weak or even negligible. Still, the effect of interaction between them on Universal dynamics cannot be ruled out. In fact, recently it has been argued by Wang et al. \cite{Wang1} that an appropriate interaction between DE and DM plays a significant role in perturbation dynamics and affect the lowest multipole of the CMB sprectum. Harko and Lobo \cite{Harko1} have also investigated the interaction between DE and DM in the context of irreversible thermodynamics of open systems with matter creation/annihilation. In this connection, it should be mentioned that the thermodynamics of irreversible processes was introduced in cosmology by Prigogine et al. \cite{Prigogine1}.

In our present work, we have investigated non-equilibrium thermodynamics of the Universe bounded by event horizon. We have examined the entropy production term by considering the non-equilibrium process of particle (DE/DM) creation/annihilation as the internal production process. Here we have considered an isentropic system consisting of two interacting dark fluids. Karami and Gaffari \cite{Karami1} have examined the validity of the generalized second law of thermodynamics (GSLT) in  a non-flat FRW universe containing interacting DE with cold DM in irreversible thermodynamical context and they have shown that GSLT is satisfied for certain range of energy-transfer constants. Here we have employed the theory of particle creation for describing interaction between DE and DM and examined the validity of GSLT for Universe bounded by the event horizon.

\section{Particle creation mechanism in the two dark species: A unified model}

Suppose a closed thermodynamical system with $N$ particles have internal energy $E$. The conservation of this internal energy gives rise to the first law of thermodynamics as \cite{Harko1}
\begin{equation}
dE=dQ-pdV,
\end{equation}
where $p$ and $V$ are the usual thermodynamic pressure and comoving volume and $dQ$ stands for the amount of heat received by the system in time $dt$. Equivalently, we have the Gibbs equation as
\begin{equation}
Tds=dq=d\left(\frac{\rho}{n}\right)+pd\left(\frac{1}{n}\right),
\end{equation}
where '$s$' is the entropy per particle, $\rho=\frac{E}{V}$ is the energy density, $n=\frac{N}{V}$ is the particle number density and $dq=\frac{dQ}{N}$ is the heat per unit particle. It should be noted that the above Gibbs equation is true for an open thermodynamical system for which the number of fluid particles is not conserved $\left(N^{\mu}_{;\mu}\neq 0\right)$. This can be expressed mathematically as
\begin{equation}
\dot{n}+\theta n=n\Gamma.
\end{equation}
Here $N^{\mu}=nu^{\mu}$ is the particle flow vector, $u^{\mu}$ is the particle four-velocity, $\theta=u^{\mu}_{;\mu}$ is the fluid expansion, $\Gamma$ stands for the rate of change of the particle number in a comoving volume $V$ and $\dot{n}=n_{,\mu}u^{\mu}$ by notation. $\Gamma$ effectively behaves as a bulk viscous pressure which causes the thermodynamics to be non-equilibrium. $\Gamma > 0$ indicates creation of particles while annihilation of particles correspond to $\Gamma <0$. 

In the present work, we consider an open thermodynamical system to be flat FRW model of the Universe with particle creation characterizing the non-equilibrium phenomenon. Suppose that the Universe consists of two dark fluids namely DM and DE. The Einstein field equations are (choosing $8\pi G=1=c$)
\begin{equation}
3H^{2}=\rho_{t}=\rho_{m}+\rho_{d}
\end{equation}
and
\begin{equation}
2\dot{H}+3H^{2}=-(p_{m}+\Pi_{m})-(p_{d}+\Pi_{d})
\end{equation}
and the energy conservation equations are given by
\begin{equation}
\dot{\rho}_{m}+3H(\rho_{m}+p_{m})=-3H\Pi_{m}
\end{equation}
and
\begin{equation}
\dot{\rho}_{d}+3H(\rho_{d}+p_{d})=-3H\Pi_{d},
\end{equation}
where $(\rho_{m},p_{m},\Pi_{m})$ are respectively energy density, thermodynamic pressure and dissipative (bulk) pressure of DM and $(\rho_{d},p_{d},\Pi_{d})$ are those of DE. The particle number conservation equations are modified as
\begin{equation}
\dot{n}_{m}+3Hn_{m}=\Gamma_{m}n_{m}
\end{equation}
and
\begin{equation}
\dot{n}_{d}+3Hn_{d}=-\Gamma_{d}n_{d},
\end{equation}
where $(n_{m},\Gamma_{m})$ are the number density and particle creation rate for DM while $(n_{d},\Gamma_{d})$ are those for DE. We have assumed that $\Gamma_{m}>0$ and $\Gamma_{d}<0$ so that DM particles are created while DE particles are destroyed, as predicted by the second law of thermodynamics \cite{Pavon1}. If for simplicity, we assume the thermodynamical system to be isentropic (i.e., entropy per particle is conserved), then the dissipative pressures and the particle creation rates are related as \cite{Pan1}
\begin{equation}
\Pi_{m}=-\frac{\Gamma_{m}}{\theta}(\rho_{m}+p_{m})
\end{equation}
and
\begin{equation}
\Pi_{d}=\frac{\Gamma_{d}}{\theta}(\rho_{d}+p_{d}).
\end{equation}
However, if the combined two-fluid is considered as a single fluid with energy density $\rho_{t}$, thermodynamic pressure $p_{t}=p_{m}+p_{d}$ and dissipative pressure $\Pi_{t}=\Pi_{m}+\Pi_{d}$, then combining (6) and (7), we have the usual conservation equation
\begin{equation}
\dot{\rho}_{t}+3H(\rho_{t}+p_{t})=-3H\Pi_{t}.
\end{equation}
Also, combining (8) and (9) we can write
\begin{equation}
\dot{n}_{t}+3Hn_{t}=\Gamma_{t}n_{t},
\end{equation}
where $n_{t}=n_{m}+n_{d}$ is the total particle number density and $\Gamma_{t}$ is the particle creation rate of the single fluid. Further, if we assume the resulting single fluid to be isentropic, then we have
\begin{equation}
\Pi_{t}=-\frac{\Gamma_{t}}{\theta}(\rho_{t}+p_{t}).
\end{equation}
Using (10) and (11) in (14), we obtain
\begin{equation}
\Gamma_{t}=\frac{\Gamma_{m}(\rho_{m}+p_{m})-\Gamma_{d}(\rho_{d}+p_{d})}{(\rho_{t}+p_{t})}.
\end{equation}
The sign of $\Gamma_{t}$ indicates whether there is a creation or an annihilation of particles of the resulting single fluid. It should be noted that if $\Pi_{m}=-\Pi_{d}<0$, then equations (6) and (7) imply interacting dark fluids with interaction term $Q=3H\Pi_{d}$ and the single fluid corresponds to a closed system as $\Gamma_{t}=0=\Pi_{t}$.

\section{Thermodynamical Analysis of Energy transfer between the dark sectors of the universe due to particle creation}

Due to the observational evidences of the present late-time acceleration, the universe is currently dominated by a dark fluid system (DM+DE). The energy conservation relations for the subsystems are given by equations (6) and (7) respectively. Due to the particle creation mechanism, there is an energy transfer between the dark species and as a result, the two subsystems have different temperatures and thermodynamics of irreversible process comes into the picture. Thus, starting from the Euler's thermodynamical relation $nTs=\rho+p$ and using the energy conservation equations (6) and (7) as well as the modified particle number conservation relations (8) and (9) for the two dark subsystems, the evolution equations for the temperatures of the individual dark fluid elements are given by
\begin{equation}
\frac{\dot{T}_{m}}{T_{m}}=-3H\left(\omega_{m}^{eff}+\frac{\Gamma_{m}}{\theta}\right)+\frac{\dot{\omega}_{m}}{1+\omega_{m}}
\end{equation}
and
\begin{equation}
\frac{\dot{T}_{d}}{T_{d}}=-3H\left(\omega_{d}^{eff}-\frac{\Gamma_{d}}{\theta}\right)+\frac{\dot{\omega}_{d}}{1+\omega_{d}},
\end{equation}
where
\begin{equation}
\omega_{m}^{eff}=\omega_{m}-\frac{\Gamma_{m}}{\theta}(1+\omega_{m})
\end{equation}
and
\begin{equation}
\omega_{d}^{eff}=\omega_{d}+\frac{\Gamma_{d}}{\theta}(1+\omega_{d}).
\end{equation}
Here, $\omega_{m}$ ($0<\omega_{m}<1$) and $\omega_{d}$ ($<-\frac{1}{3}$) are the equations of state parameters for DM and DE respectively and $T_{m}, T_{d}$ are respectively the temperature of the two dark components. As for simplicity, the thermodynamical process is assumed to be adiabatic, so the effective bulk pressures $\Pi_{m}$ and $\Pi_{d}$ are respectively related to the corresponding particle creation rates by Eqs. (10) and (11). Now integrating Eqs. (16) and (17), we have the temperatures of the two dark sectors as
\begin{equation}
T_{m}=T_{0}(1+\omega_{m})exp\left[-3\int_{a_{0}}^{a}\left(\omega_{m}^{eff}+\frac{\Gamma_{m}}{\theta}\right)\frac{da}{a}\right]
\end{equation}
and
\begin{equation}
T_{d}=T_{0}(1+\omega_{d})exp\left[-3\int_{a_{0}}^{a}\left(\omega_{d}^{eff}-\frac{\Gamma_{d}}{\theta}\right)\frac{da}{a}\right],
\end{equation}
where $T_{0}$ is the common temperature of the two subsystems in equilibrium configuration and $a_{0}$ is the value of the scale factor in the equilibrium state. It should be noted that at the very early stages of the evolution of the universe, $T_{m}> T_{d}$ and with the evolution of the universe, both the subsystems approach the equilibrium configuration ($a=a_{0}$) with common temperature $T_{0}$. Then for $a>a_{0}$,the thermodynamical equilibrium is violated due to the continuous flow of DE to DM and as a result we have $T_{m}<T_{0}<T_{d}$. In the present case, we are considering universe bounded by the event horizon as an isolated thermodynamical system and so in the equilibrium state the temperature $T_{0}$ may be considered as the (modified) Hawking temperature \cite{Chakraborty1, Saha1} on the event horizon, i.e., $T_{0}=\frac{H^{2}R_{E}}{2\pi}|_{a=a_0}$, where $R_{E}$ is the radius of the event horizon for the FRW model. Further, for thermodynamical analysis, we denote '$S_{m}$' and '$S_{d}$' as the entropies of the two dark sectors of the present isolated system while $S_{E}$ stands for the entropy of the bounding event horizon. Then, from the first law (i.e., Clausius relation) for individual subsystems, we have
\begin{equation}
T_{m}\frac{dS_{m}}{dt} =\frac{dQ_{m}}{dt}=\frac{dE_{m}}{dt}+p_{m}\frac{dV}{dt}
\end{equation}
and
\begin{equation}
T_{d}\frac{dS_{d}}{dt} =\frac{dQ_{d}}{dt}=\frac{dE_{d}}{dt}+p_{d}\frac{dV}{dt}.
\end{equation}
In the above expressions, $E_{m}=\rho_{m}V$ and $E_{d}=\rho_{d}V$ are the energies of the two subsystems while $Q_{m}$ and $Q_{d}$ are the amount of heat of the two dark sectors and $V=\frac{4}{3}\pi R_{E}^{3}$ is the volume of the universe bounded by the two subsystems. Due to the isolated nature of the whole system, the heat flow across the horizon ($Q_{h}$) is balanced by the heat flow through the two dark components, i.e.,
\begin{equation}
\dot{Q}_{h}=-\left(\dot{Q}_{m}+\dot{Q}_{d}\right).
\end{equation}
However, in view of non-equilibrium thermodynamical prescription one has to take care of the contributions from irreversible fluxes of energy transfer in the expression for the total entropy variation as \cite{Karami1, Zhou1}
\begin{equation}
\frac{dS_{T}}{dt}=\frac{dS_{m}}{dt}+\frac{dS_{d}}{dt}+\frac{dS_{E}}{dt}-A_{d}\dot{Q}_{d}\ddot{Q}_{d}-A_{h}\dot{Q}_{h}\ddot{Q}_{h},
\end{equation}
where $A_{d}$ and $A_{h}$ are the energy transfer constants between DE and DM within the universe and between the universe and the horizon respectively.

Here our universe is described by a 4$-$dimensional flat FRW model bounded by an event horizon. Normally in FRW spacetime, event horizon does not exist. It is relevant only when the universe is in an accelerating phase. So in the perspective of the present accelerating phase of the universe, it is reasonable to consider universe bounded by the event horizon as an open thermodynamical system. For event horizon, the area radius has the expression
\begin{equation}
R_{E}=a\int_{t}^{\infty}\frac{dt}{a}
\end{equation}
(Note that the above improper integral exists when the universe is in an accelerating phase) and variation of $R_{E}$ with respect to cosmic time $t$ gives
\begin{equation}
\dot{R}_{E}=HR_{E}-1.
\end{equation}
Also, let us consider Bekenstein entropy-area relation on the surface area of the event horizon, i.e.,
\begin{equation}
S_{E}=\pi R_{E}^{2}.
\end{equation}
Now we shall give two examples of DE model and discuss the above process in respective scenarios.

\subsection{Interacting dark energy with constant equation of state}

Here we shall consider DE as a perfect fluid with an equation of state $p_{d}=\omega_{d}\rho_{d}$ where $\omega_{d}$ is a constant such that $\omega_{d}<-\frac{1}{3}$ and DM to be in the form of pressureless dust, i.e., $p_{m}=0$. We choose dissipative pressures of DM and DE as
\begin{equation}
-\Pi_{m}=\Pi_{d}=\lambda \rho_{d}
\end{equation}
so that the particle creation rates for DM and DE become (from Eq. (10))
\begin{equation}
\Gamma_{m}=3H\lambda\frac{\rho_{d}}{\rho_{m}}~,~~\Gamma_{d}=-\frac{3H\lambda}{1+\omega_{d}}
\end{equation}
respectively. The temperatures of these two dark sectors read as 
\begin{equation}
T_{m}=T_{0}~~~and~~~T_{d}=T_{0}(1+\omega_{d})\left(\frac{a}{a_{0}}\right)^{-3\left(\lambda+\frac{\lambda} {1+\omega_{d}}+\omega_{d}\right)}.
\end{equation}
Then, one gets (assuming $8\pi=1=G$)
\begin{equation}
\dot{Q}_{m}=\frac{3}{2}(HR_E)^2(\lambda \Omega _d HR_E+\Omega _d -1),
\end{equation}
\begin{equation}
\dot{Q}_{d}=-\frac{3}{2}(HR_{E})^{2}\Omega_{d}(\lambda HR_{E}+\omega_{d} +1),
\end{equation}
and
\begin{equation}
\dot{Q}_{h}=\frac{3}{2}(HR_{E})^{2}(1+\omega_{d}\Omega_{d}).
\end{equation}
Differentiating (33) and (34), we have 
\begin{equation}
\ddot{Q}_d=\frac{3}{2}H^2 R_E \Omega _d \left[\lambda HR_E(1+qHR_E)+(\lambda HR_E+\omega _d+1)\left\lbrace 2(1+qHR_E)+3HR_E(\omega _d+\omega _d \Omega _d+\lambda +2) \right\rbrace\right]
\end{equation}
and
\begin{equation}
\ddot{Q}_{h}=-\frac{3}{2}H^2 R_E\left[3HR_E \omega _d \Omega _d (\omega _d+\omega _d \Omega _d+\lambda +2)+2(1+\omega _d \Omega _d)(1+qHR_E)\right]
\end{equation}
respectively, where the deceleration parameter $q$ can be evaluated as
\begin{equation}
q=-\frac{\dot{H}}{H^2}-1=\frac{1}{2}(1+3\omega _d \Omega _d).
\end{equation}
Also, 
\begin{equation}
\dot{S}_{E}=\frac{1}{4}R_E (HR_E-1).
\end{equation}
Thus, from Eq. (25), we get
\begin{eqnarray}
\frac{dS_{T}^{E}}{dt}&=&\frac{3}{2T_0} (HR_E)^2(\lambda \Omega _d HR_E+\Omega _d -1)-\frac{3}{2T_0 (1+\omega _d)}\left(\frac{a}{a_0}\right)^{3(\lambda +\omega _d+\frac{\lambda}{(1+\omega _d)})}(HR_E)^2 \Omega _d(\lambda HR_E+\omega _d+1) \nonumber \\
&+& \frac{1}{4} R_E(HR_E-1)+\frac{9}{4}A_dH^4 R_{E}^{3}{\Omega _d}^2 \left(\lambda HR_E+\omega _d+1\right)\Big[\lambda H^2R_E (1+qHR_E)+(\lambda HR_E+\omega _d+1) \nonumber \\
&\times & \left\lbrace 2(1+qHR_E)+3HR_E(\omega _d+\omega _d \Omega _d+\lambda +2)\right\rbrace \Big]+\frac{9}{4}A_h H^4 R_{E}^{3} (1+\omega _d \Omega _d)\Big[3HR_E\omega _d \Omega _d \nonumber \\
&\times & (\omega _d+\omega _d \Omega _d+\lambda +2)+2(1+\omega _d \Omega _d)(1+qHR_E)\Big].
\end{eqnarray}
Due to a complicated expression of $\frac{dS_{T}^{E}}{dt}$, its evolution against $\lambda$ has been plotted in Fig. 1 for different choices of $w_d$ in quintessence, phantom and normal fluid eras. From the figure, we see that if the DE is either not close to $\Lambda$CDM or in phantom domain, then GSLT holds for all values of $\lambda$ while for DE in phantom domain, GSLT does not hold for very small values of $\lambda$. Also, for $w_d=-1.4$, the total entropy variation shows some peculiar fluctuations.

\begin{figure}
\begin{minipage}{0.4\textwidth}
\includegraphics[width=1.0\linewidth]{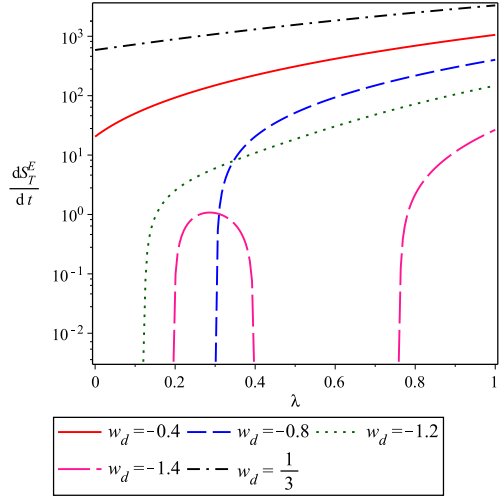}
\end{minipage}
\caption{The plots show the evolution of $\frac{dS_{T}^{E}}{dt}$ against $\lambda$ for different values of $\omega _d$ in Quintesssence ($\omega _d=-0.4, -0.8$), Phantom ($\omega _d=-1.2, -1.4$) and Normal fluid ($\omega _d=\frac{1}{3}$) eras.}
\end{figure}

\subsection{Interacting Holographic Dark Energy Model}

The holographic dark energy density satisfies the relation $\rho_{d}=\frac{3c^{2}}{L^{2}}$ \cite{Hsu1, Li1}, where $c$ is a constant to be determined from observations and $L$ is an IR cut-off in units of $8\pi =1=G$. Li \cite{Li1} has shown that if we choose $L$ as the radius of the future event horizon ($R_{E}$), only then we can obtain the correct equation of state of DE and the present accelerating universe. Moreover, interaction models have been shown to support \cite{Das1, Amendola1} the observationally measured phantom equation of state and are also favored by observational data from the cosmic microwave background (CMB) \cite{Olivares1} and the matter distribution at large scales \cite{Olivares2}. Thus, with the choice $L=R_{E}$, we can write
\begin{equation}
R_{E}=\frac{c}{H\sqrt{\Omega_{d}}},
\end{equation}
where $\Omega_{d}=\frac{\rho_{d}}{3H^{2}}$ is the DE density parameter. As in case A, here also we choose the dissipative pressures as $-\Pi_{m}=\Pi_{d}=\lambda \rho =\lambda \rho_d$, $\lambda$ being a small positive dimensionless parameter. Consequently, the particle creation rates for DM and DE take the forms $\Gamma_{m}=3H\lambda\frac{\rho_{m}}{\rho_{d}}$ and $\Gamma_{d}=-\frac{3H\lambda}{1+\omega_{d}}$ respectively.
 
The density parameter evolves as \cite{Wang1, Wang2}
\begin{equation}
\Omega_{d}'=\Omega_{d}\left[(1-\Omega_{d})\left(1+\frac{2\sqrt{\Omega_{d}}}{c}\right)-3\lambda \Omega _d\right ]
\end{equation}
while the variable equation of state for holographic DE is given by \cite{Wang1}
\begin{equation}
\omega_{d}=-\frac{1}{3}-\frac{2}{3c}\sqrt{\Omega_{d}}-\lambda.
\end{equation}
Note that in Eq. (41), the prime refers to derivative with respect to $x=ln~a$. Here the temperatures of DM and DE can be written as
\begin{equation}
T_{m}=T_{0}~~~and~~~T_{d}=T_{0}(1+\omega _d)exp \left[-3\int _{a_0}^{a} \left(\omega _d+\lambda \frac{(2+\omega _d)}{(1+\omega _d)}\right)\frac{da}{a}\right].
\end{equation}
So, we now obtain
\begin{equation}
\dot{Q}_{m}=\frac{3c^2}{2} \left(\lambda \frac{c}{\sqrt{\Omega _d}}-\frac{1}{\Omega _d} +1 \right),
\end{equation}
\begin{equation}
\dot{Q}_{d}=-\frac{3c^2}{2} \left(\lambda \frac{c}{\sqrt{\Omega _d}}+\omega_{d} +1 \right),
\end{equation}
and
\begin{equation}
\dot{Q}_{h}=\frac{3c^2}{2} \left(\frac{1}{\Omega _d}+\omega _d\right).
\end{equation}
Differentiating (45) and (46), we have 
\begin{eqnarray}
\ddot{Q}_d&=&\frac{3}{2}Hc\sqrt{\Omega _d}\Big[\frac{c}{\sqrt{\Omega _d}}\left\lbrace \lambda (1+q\frac{c}{\sqrt{\Omega _d}})+\frac{\sqrt{\Omega _d}}{3c}\left\lbrace (1-\Omega _d)\left(1+\frac{2}{c}\sqrt{\Omega _d}\right)-3\lambda \Omega _d \right\rbrace\right\rbrace +(\lambda \frac{c}{\sqrt{\Omega _d}} \nonumber \\
&+& \omega _d+1)\left\lbrace \frac{2c}{\sqrt{\Omega _d}}(1+q)-2\left(\frac{c}{\sqrt{\Omega _d}}-1\right)-\frac{c}{\sqrt{\Omega _d}} \left\lbrace (1-\Omega _d)\left(1+\frac{2}{c}\sqrt{\Omega _d}\right)-3\lambda \Omega _d \right\rbrace \right\rbrace \Big]
\end{eqnarray}
and
\begin{eqnarray}
\ddot{Q}_{h}&=&\frac{3}{2}H\frac{c}{\sqrt{\Omega _d}}\Big[c\sqrt{\Omega _d}\left\lbrace (1-\Omega _d)\left(1+\frac{2}{c}\sqrt{\Omega _d}\right)-3\lambda \Omega _d \right\rbrace \left(\omega _d-\frac{1}{3c}\sqrt{\Omega _d}\right)-2(1+\omega _d \Omega _d)(1 \nonumber \\
&+& qHR_E)\Big].
\end{eqnarray}
Also, 
\begin{equation}
\dot{S}_E=\frac{1}{4}\frac{c}{H\sqrt{\Omega _d}}\left(\frac{c}{\sqrt{\Omega _d}}-1\right).
\end{equation}
Then, from Eq. (25), we have
\begin{eqnarray}
\frac{dS_{T}^{E}}{dt}&=&\frac{3c^2}{2T_0} \left(\lambda \frac{c}{\sqrt{\Omega _d}}-\frac{1}{\Omega _d} +1 \right)-\frac{3c^2}{2T_0(1+\omega _d)} exp \left[3\int _{a_0}^{a} \left(\omega _d+\lambda \frac{(2+\omega _d)}{(1+\omega _d)}\right)\frac{da}{a}\right]\left(\lambda \frac{c}{\sqrt{\Omega _d}}+\omega_{d} +1 \right) 
\nonumber \\
&+& \frac{1}{4}\frac{c}{H\sqrt{\Omega _d}}\left(\frac{c}{\sqrt{\Omega _d}}-1\right)+\frac{9}{4}A_d H\left(\frac{c}{\sqrt{\Omega _d}}\right)^3 \Omega _{d}^{2} \left(\lambda \frac{c}{\sqrt{\Omega _d}}+\omega _d+1 \right) \Big[\frac{c}{\sqrt{\Omega _d}} \nonumber \\
&\times &\left\lbrace \lambda (1+q\frac{c}{\sqrt{\Omega _d}})+\frac{\sqrt{\Omega _d}}{3c} \left\lbrace (1-\Omega _d)\left(1+\frac{2}{c}\sqrt{\Omega _d}\right)-3\lambda \Omega _d \right\rbrace\right\rbrace +(\lambda \frac{c}{\sqrt{\Omega _d}}+\omega _d+1) \nonumber \\
&\times & \left\lbrace \frac{2c}{\sqrt{\Omega _d}}(1+q)-2\left(\frac{c}{\sqrt{\Omega _d}}-1\right)-\frac{c}{\sqrt{\Omega _d}} \left\lbrace (1-\Omega _d)\left(1+\frac{2}{c}\sqrt{\Omega _d}\right)-3\lambda \Omega _d \right\rbrace \right\rbrace \Big]-\frac{9}{4}A_h H\left(\frac{c}{\sqrt{\Omega _d}}\right)^3 \nonumber \\
&\times & (1+\omega _d \Omega _d)\Big[c\sqrt{\Omega _d}\left\lbrace (1-\Omega _d)\left(1+\frac{2}{c}\sqrt{\Omega _d}\right)-3\lambda \Omega _d \right\rbrace \left(\omega _d-\frac{1}{3c}\sqrt{\Omega _d}\right)-2(1+\omega _d \Omega _d) \nonumber \\
&\times & (1+qHR_E)\Big].
\end{eqnarray}

\begin{figure}
\begin{minipage}{0.4\textwidth}
\includegraphics[width=1.0\linewidth]{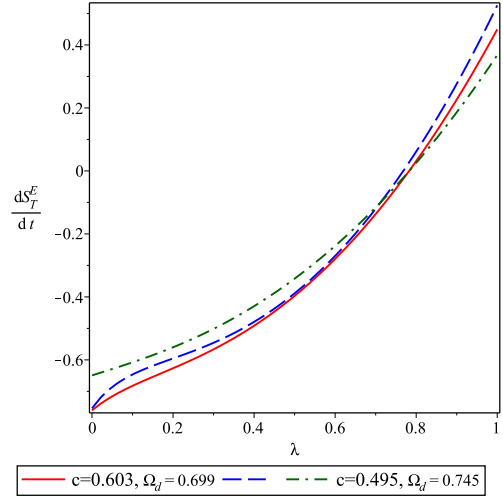}
\end{minipage}
\caption{The plot shows the evolution of $\frac{dS_{T}^{E}}{dt}$ against $\lambda$ for three {\it Planck} data sets.}
\end{figure}

Here again due to complicated mathematical expression for $\frac{dS_{T}^{E}}{dt}$, we have shown its evolution against the parameter $\lambda$ graphically for three {\it Planck} data sets \cite{Li2} providing values for $c$ and $\Omega _d$. It is evident that GSLT holds for $\lambda \gtrsim 0.8$.

\section{Summary and Conclusions}

The present work deals with non-equilibrium thermodynamics in the frame work of particle creation mechanism. At first, it has been shown that two fluids with different particle creation rates can be considered as a single fluid. However, if the dissipative pressure for the two fluids are identical in magnitude but different in sign then the resulting single fluid has no particle creation rate and as a result it corresponds to a close system. Then considering the two dark species as DM (in the form of dust) and DE (as perfect fluid with a constant equation of state in one subsection and as holographic DE in another subsection), firstly we have shown that DM has a higher temperature at the initial stages and then it gradually decreases and attains a equilibrium era and finally the DE temperature dominates. Although the choice of the dissipative pressures (as in Eq. (29)) is purely phenomenological but from the mathematical point of view, it is easily solvable for both the choices of DE. Also physically we can argue as follows: As $\rho _d \propto H^2$, so dissipative pressure is of the order of $H^2$ and it is consistent with the choice of Barrow \cite{Barrow1}. Subsequently, in both the cases, analytic expressions for the total entropy variations have been determined. Due to complicated expressions we have graphically examined the validity of GSLT. In case of constant equation of state, GSLT holds for almost all values of $\lambda$ (except for very small $\lambda$ for $\omega _{d} \lesssim -1$). For HDE model, we have chosen three {\it Planck} data sets for observed values of $c$ and $\Omega _d$ and it is observed that GSLT holds for all the three {\it Planck} data sets for $\lambda \gtrsim 0.8$.

\section*{Acknowledgments}
All the authors are thankful to IUCAA for research facilities as a part of the work was done there during a visit. The author S.S. is also thankful to UGC-BSR Programme of Jadavpur University for providing research fellowship. The author S.C. is thankful to the UGC-DRS Programme in the Department of Mathematics, Jadavpur University.

\end{document}